\documentclass[letterpaper,12pt]{article}

 \title{Backtesting forecast accuracy}
 \author{Arturo Erdely}
 \date{\small{Facultad de Estudios Superiores Acatl\'an \\
              Universidad Nacional Aut\'onoma de M\'exico \\
							\texttt{arturo.erdely@comunidad.unam.mx}\\}}

 \usepackage{amsthm}
 \usepackage{amssymb}
 \usepackage{amsmath}
 \usepackage{enumerate}
 \usepackage{graphicx}
 
\begin{document}
 
\maketitle

\begin{abstract}
\noindent A statistical test based on the geometric mean is proposed to determine if a predictive model should be rejected or not, when the quantity of interest is a strictly positive continuous random variable. A simulation study is performed to compare test power performance against an alternative procedure, and an application to insurance claims reserving is illustrated.
\end{abstract}


\section{Introduction}

From Diebold and Mariano (2001):
\begin{quote}
  Prediction is of fundamental importance in all of the sciences, including economics. Forecast accuracy is of obvious importance to users of forecasts because forecasts are used to guide decisions. Forecast accuracy is also of obvious importance to producers of forecasts, whose reputations (and fortunes) rise and fall with forecast accuracy [\ldots] Predictive performance and model adequacy are inextricably linked---predictive failure implies model inadequacy.
\end{quote}

\noindent Several measures of forecast accuracy have been proposed, mainly for the purpose of comparing two or more forecasting competing methods, see for example Hyndman and Koehler (2006) or Shcherbakov \textit{et al.}(2013), and statistical tests have been proposed for the null hypothesis of equal forecast accuracy of such competing methods, see Diebold and Mariano (2001). But if the objective is to evaluate a single forecasting method, the value itself reported by a measure of forecast accuracy does not allow us to assess how good/bad such method is, since it is made for comparison purposes, it is a \textit{relative} rather than an \textit{absolute} measure for the forecasting quality of the predictive model in question.

\bigskip

\noindent In the present work a statistical test is proposed to decide whether a single forecasting method should be rejected or not as accurate, based on the \textit{geometric mean} of the ratios of observed/forecasted values, in the particular case when all the involved quantities are strictly positive. A simulation study is performed to compare the statistical power of the proposed \textbf{accuracy test} versus a \textit{binomial test}, and an application to insurance claims reserving is illustrated.

\section{Accuracy test}

\noindent Consider the case when all the observed ($s_i$) and forecasted ($r_i$) values are strictly positive, for $i=1,\ldots,n.$ If we calculate the ratios $x_i=s_i/r_i$ then $x_i=1$ means a perfect forecast, $x_i<1$ implies overestimation, and $x_i>1$ underestimation by forecast $i.$ We will assume that in $\{x_i:i=1\ldots,n\}$ there are no repeated values and that these values may be considered as an observed random sample from a strictly positive and continuous random variable $X$ with unknown probability density function (pdf) $f_X\,.$ 

\bigskip

\begin{figure}
  \centerline{
  \includegraphics[width=13cm, keepaspectratio]{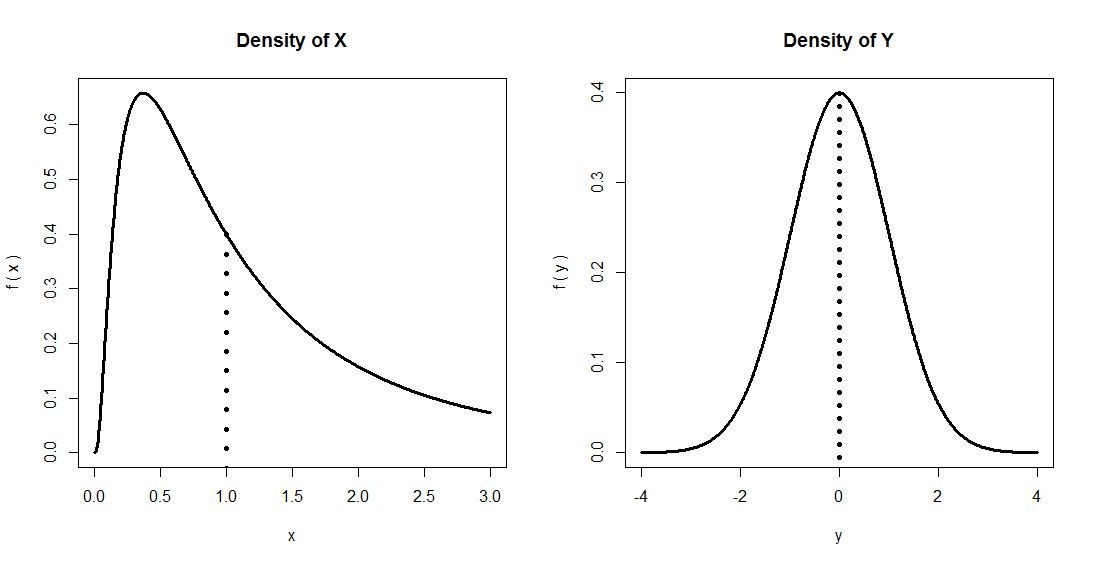}}
  \caption{Left: Lognormal pdf with parameters $\mu=0$ and $\theta=1.$ Right: Normal pdf with $\mu=0$ and $\theta=1.$}
  \label{YlogX}
\end{figure}

\noindent The hypothesis of interest is whether or not $f_X$ is centered around $1$ (the value that represents perfect forecasts) under a certain \textit{measure of central location}. If this is the case, the pdf $f_X$ may look something like in Fig.\ref{YlogX}(left), and if we make the transformation $Y:=\log X$ the resulting pdf $f_Y$ may look something like in Figure \ref{YlogX}(right), since the $\log$ function maps $0<X<1$ values to the open interval $\,]-\infty,0\,[\,$ and $X\geq 1$ values are mapped to $[\,0,+\infty[\,.$ 

\bigskip

\noindent If we define the transformed values $y_i=\log x_i$ $(i=1,\ldots,n)$ these may be considered as an observed random sample from a random variable $Y=\log X.$ If a statistical test applied to the $y_i$ values does not reject the null hypothesis of normality this would be equivalent to not rejecting that the $x_i$ values are an observed random sample from a LogNormal distribution, that is, if $Y$ is a Normal$(\mu,\theta)$ random variable then $X=e^Y$ would be a LogNormal$(\mu,\theta)$ random variable.

\bigskip

\noindent As stated in Johnson \textit{et al.}(1994), even without the lognormality assumption, if $X_1,\ldots,X_n$ are independent and continuous positive random variables, and $W_n:=\prod_{i\,=\,1}^n X_i$ then $\log W_n=\sum_{i\,=\,1}^n\log X_i$ and if the independent random variables $Y_i=\log X_i$ are such that a central limit type of result applies, then the  standardized distribution of $\log W_n$ would tend to a standard Normal distribution as $n$ tends to infinity, and the limiting distribution of $W_n$ would then be LogNormal.

\bigskip

\noindent Closely related to the random variable $W_n:=\prod_{i\,=\,1}^n X_i$ is the concept of \textit{sample geometric mean} $\widetilde{X}:=(W_n)^{1/n}$ and then $\log\widetilde{X}=\frac{1}{n}\sum_{i\,=\,1}^n\log X_i=\overline{Y}$ (the sample mean of $Y_1,\ldots,Y_n$), so that $\widetilde{X}$ may also have either an exact or limiting LogNormal distribution, according to the above arguments. As defined in Kotz \textit{et al.}(2006) for a random variable $X$, a parameter analogous to the sample geometric mean is $\mathbb{GM}(X) := \exp(\mathbb{E}[\log X])$ when $\mathbb{E}[|\log X|] < \infty.$ In the following, $\mu$ will denote $\mathbb{E}[\log X]$ so that $\mathbb{GM}(X) = e^{\,\mu}.$ Therefore, if $X$ is distributed LogNormal$(\mu,\theta)$ then $\mathbb{GM}(X)=e^{\,\mu}$ since $\mathbb{E}[\log X]=\mu\,.$ By the way, for the lognormal distribution the geometric mean is equal to its median.

\bigskip

\noindent Assuming we have a random sample of observed/forecasted ratios $X_1,\ldots,X_n$ distributed as\linebreak LogNormal$(\mu,\theta)$ we have the following equivalent ways of expressing the null hypothesis of interest:
\begin{equation}\label{Ho}
\mathcal{H}_0:\mathbb{GM}(X)=1\qquad\equiv\qquad\mathcal{H}_0: e^{\,\mu}=1\qquad\equiv\qquad\mathcal{H}_0:\mu=0
\end{equation}

\noindent Under $\mathcal{H}_0$ the transformation $Y_i:=\log X_i$ leads to a random sample from a Normal$(\mu,\theta)$ distribution, therefore we may use the standard $t-$test of size $0<\alpha<1$ for $\mathcal{H}_0:\mu=0$ with unknown variance $\theta>0,$ see for example Mood \textit{et al.}(1974), rejecting $\mathcal{H}_0$ whenever $|T|>k_{\alpha}\,,$ where $T=\overline{Y}\sqrt{n}/S_Y,$ $S_Y^2=\frac{1}{n-1}\sum_{i\,=\,1}^n(Y_i-\overline{Y})^2$ and $k_\alpha$ is the $1-\alpha/2$ quantile of a $t$ distribution with $n-1$ degrees of freedom.

\bigskip

\noindent Finally, to validate the lognormality assumption of the the observed/forecasted ratios $X_1,\ldots,X_n$ we may apply a normality test to $\log X_1,\ldots,\log X_n$ such as the Shapiro-Wilk (1965) test since it is a more powerful test for normality than the tests by Anderson-Darling, Lilliefors and Kolmogorov-Smirnov, according to a power study by Razali and Wah (2011).

\section{Power study}

Define the random variables $R\sim\text{LogNormal}(\mu_R,\theta_R)$ and $S\sim\text{LogNormal}(\mu_S,\theta_S),$ then $\mathbb{GM}(R)=e^{\mu_R}$ and $\mathbb{GM}(S)=e^{\mu_S}.$ If we define the ratio random variable $X:=S/R$ then $\log X = \log S - \log R,$ and since $\log S\sim\text{Normal}(\mu_S,\theta_S)$ and $\log R\sim\text{Normal}(\mu_R,\theta_R)$ then $\log X$ is normally distributed with mean $\mu_S-\mu_R$ and variance $\theta_S+\theta_R-2\rho\sqrt{\theta_S\theta_R}$ (where $\rho$ stands for linear correlation coefficient of $\log S$ and $\log R$, which is also referred to as \textit{log-correlation}) and therefore the ratio $X$ has LogNormal distribution with parameters equal to the previous mean and variance of $\log X.$ This implies that $\mathbb{GM}(X)=e^{\mu_S-\mu_R}=\mathbb{GM}(S)/\mathbb{GM}(R).$ Just as an observation, this is not generally true for the usual mean: $\mathbb{E}(X)$ is not equal to $\mathbb{E}(S)/\mathbb{E}(R)$ unless $\theta_R=\rho^2\theta_S\,$ (under lognormality).

\bigskip

\noindent Under the above assumptions, the null hypothesis of interest (\ref{Ho}) would be equivalent to $\mathcal{H}_0:\mathbb{GM}(S)=\mathbb{GM}(R)$ or $\mathcal{H}_0:\mu_S=\mu_R\,.$ Let $\beta>-1$ such that $\mathbb{GM}(S)=(1+\beta)\mathbb{GM}(R)\,,$ then $\mu_S=\log(1+\beta)+\mu_R$ and (\ref{Ho}) will also be equivalent to $\mathcal{H}_0:\beta=0\,.$ In a power study for the accuracy test proposed in the previous section, it would be expected a low probability of rejection whenever $\beta=0$ (type I error) and higher probabilities of rejection as $\beta$ gets closer either to $-1$ or to $+\infty.$

\bigskip

\noindent If in addition to $\mu_S=\mu_R$ we have that $\theta_S=\theta_R$ then the ratio $X\sim\text{LogNormal}(0,2\theta_R(1-\rho)),$ that is, for fixed $\theta_R,$ the lower the linear correlation between $\log S$ and $\log R$ the larger the variability of $X$ and vice versa, as expected. So in this power study several scenarios are considered with different values for both $\beta$ and $\rho,$ for fixed $\theta_R=1$ and sample sizes $n\in\{20, 100\}.$

\bigskip

\noindent The accuracy test proposed in the previous section is compared to the usual backtesting technique based on a binomial test for the number of exceptions (or violations) of a given $\text{VaR}_\varepsilon$ (Value-at-Risk of level $\varepsilon$) as in (for example) McNeil \textit{et al.}(2015) with $\varepsilon=1/2,$ since under $\mathcal{H}_0$ it would be expected a balance between forecasts above and below the observed values, and therefore too many forecasts above or below observed values should lead to the rejection of the forecasting method. 

\bigskip

\noindent To be more specific, as a summary of both the proposed accuracy test and the binomial test of level $0<\alpha<1$, given the forecasts $r_1,\ldots,r_n$ and the corresponding observed values $s_1,\ldots,s_n\,,$ calculate the ratios $x_i=s_i/r_i$ and:

\medskip

\noindent \underline{\textbf{Accuracy test}}
\begin{enumerate}
  \item Calculate $y_1,\ldots,y_n$ by $y_i=\log x_i\,.$
	\item Apply Shapiro-Wilk normality test to the set $\{y_i:i=1,\ldots,n\}.$ If normality is rejected, the test should not be used; otherwise, continue.
	\item Apply a $t-$test for $\mathcal{H}_0:\mu=0$ to the set $\{y_i:i=1,\ldots,n\}.$ If the null hypothesis is rejected (\textit{p-value} $\leq\alpha$) the forecast method is considered inaccurate.
\end{enumerate}

\noindent \underline{\textbf{Binomial test}}
\begin{enumerate}
  \item Calculate $b$ the number of $x_i$ values greater than $1.$
	\item Calculate (two-sided test) \textit{p-value} with $B\sim\text{Binomial}(n,\frac{1}{2}),$ that is, if $b>\frac{n}{2}$ then \textit{p-value} is $2\hspace{0.3mm}\mathbb{P}(B\geq b)=\frac{1}{2^{n-1}}\sum_{x\,=\,b}^n\binom{n}{x},$ and if $b<\frac{n}{2}$ then \textit{p-value} is $2\hspace{0.3mm}\mathbb{P}(B\leq b)=\frac{1}{2^{n-1}}\sum_{x\,=\,0}^b\binom{n}{x}.$ In case $b=\frac{n}{2}$ that is exactly the center of the distribution, that is $\mathbb{E}(B)=\frac{n}{2}\,,$ and therefore the \textit{p-value} should be defined as $1.$ 
	\item If \textit{p-value} $\leq\alpha$ the forecast method is considered inaccurate.
\end{enumerate}

\begin{figure}
  \centerline{
  \includegraphics[width=16cm, keepaspectratio]{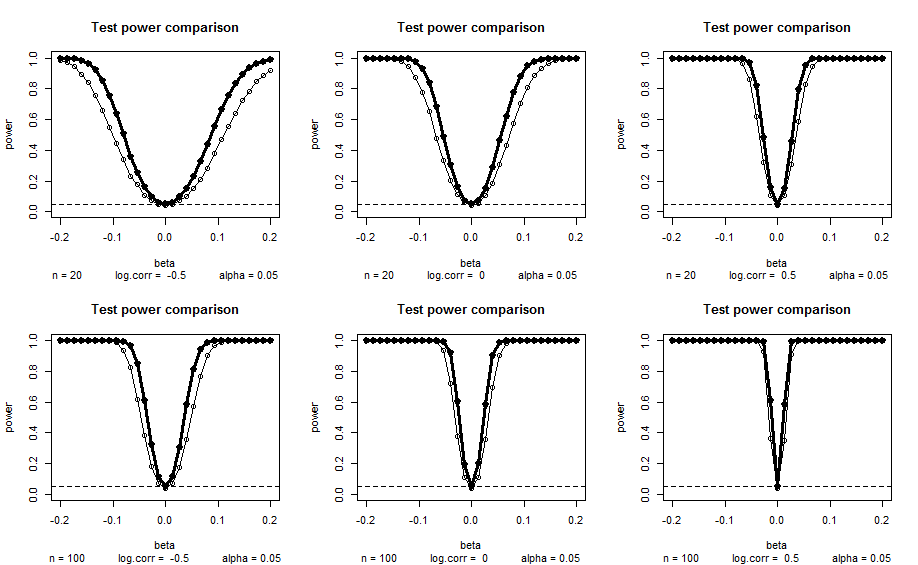}}
  \caption{Estimated power functions for the proposed accuracy test (thick line) versus the binomial test (thin line), for $-0{.}2<\beta<+0{.}2,$ log-correlations $\{-0{.}5, 0, +0{.}5\}$ and sample sizes $n\in\{20, 100\},$ with $\alpha = 0{.}05$ test level.}
  \label{powerLN}
\end{figure}

\noindent $10,000$ simulations were used to estimate each point of the power functions and the results obtained are summarized in Figure \ref{powerLN} where it is clear that in all cases the proposed accuracy test (thick line) is uniformly more powerful than the binomial test (thin line), which should not be a surprise mainly for two reasons: first, the lognormality assumption is guaranteed; second, the proposed accuracy test is based on the sample geometric mean, which is a \textit{sufficient statistic} for the theoretical geometric mean, while the binomial test is based on counts of observed above/below forecasts, which is certainly not a sufficient statistic for the parameter of interest.

\bigskip

\noindent Now let $R$ and $S$ be continuous strictly positive \textbf{non LogNormal} random variables, such that $\mathbb{E}[|\log R|]<\infty$ and $\mathbb{E}[|\log S|]<\infty.$ Define $\mu_R:=\mathbb{E}[\log R]$ and $\mu_S:=\mathbb{E}[\log S],$ then $\mathbb{GM}(R)=e^{\mu_R}$ and $\mathbb{GM}(S)=e^{\mu_S}.$ Define the ratio random variable $X:=S/R$ then $\mathbb{E}[|\log X|]$ is finite since $$\mathbb{E}[|\log X|]=\mathbb{E}[|\log S - \log R|]\leq\mathbb{E}[|\log S|+|\log R|]=\mathbb{E}[|\log S|]+\mathbb{E}[|\log R|] < \infty$$ and so $\mathbb{GM}(X)=\exp(\mathbb{E}[\log X])=e^{\mu_S-\mu_R}=\mathbb{GM}(S)/\mathbb{GM}(R).$ Under these assumptions, the null hypothesis of interest (\ref{Ho}) would be (again) equivalent to $\mathcal{H}_0:\mathbb{GM}(S)=\mathbb{GM}(R)$ or $\mathcal{H}_0:\mu_S=\mu_R\,.$ Let $\beta>-1$ such that $\mathbb{GM}(S)=(1+\beta)\mathbb{GM}(R)\,,$ then $\mu_S=\log(1+\beta)+\mu_R$ and (\ref{Ho}) will also be equivalent to $\mathcal{H}_0:\beta=0\,.$ In a power study for the accuracy test proposed in the previous section, it would be expected a low probability of rejection whenever $\beta=0$ (type I error) and higher probabilities of rejection as $\beta$ gets closer either to $-1$ or to $+\infty.$

\bigskip

\noindent Lets analyze, for example, the case when $R\sim\text{Gamma}(a_R, b_R)$ and $S\sim\text{Gamma}(a_S, b_S)$ where the pdf for a Gamma random variable $Z$ with parameters $a>0$ and $b>0$ is given by $$f_Z(z\,|\,a,b) \,=\, \frac{b^a}{\Gamma(a)}\,z^{a-1}e^{-bz}\,,\quad z > 0$$ with $\mathbb{E}(Z)=\frac{a}{b}\,,$ $\mathbb{V}(Z)=\frac{a}{b^2}\,,$ $\mathbb{E}(\log Z)=\psi(a)-\log b\,,$ and so $\mathbb{GM}(Z)=\frac{1}{b}e^{\psi(a)}\,,$ where $\psi$ is the \textit{digamma function} given by $\psi(a):=d\log\Gamma(a)/da=\Gamma'(a)/\Gamma(a)\,.$ We get then the following equivalences:
\begin{equation}\label{gammaGM}
\mathbb{GM}(Z)\,=\,1\qquad\Leftrightarrow\qquad e^{\psi(a)}\,=\,b\qquad\Leftrightarrow\qquad \psi(a)=\log b
\end{equation}

\noindent Given $Z\sim\text{Gamma}(a,b)$ if we define the transformation $Y:=\log Z$ then $Y$ is a continuous random variable with support the whole real line $\mathbb{R}$ and with a standard probability transformation procedure the following is the resulting pdf of $Y:$ 
\begin{equation}\label{logGamma}
  f_Y(y\,|\,a,b)\,=\,\frac{b^a}{\Gamma(a)}\,e^{ay-be^y}\,,\qquad y\in\mathbb{R}
\end{equation}
and clearly $\log Z$ is not Normal when $Z$ is Gamma (in fact, the probability distribution of $\log Z$ is known as \textit{LogGamma} distribution.) If $R\sim\text{Gamma}(a_R, b_R)$ and $S\sim\text{Gamma}(a_S, b_S)$ and $X:=S/R$ then $\log X=\log S-\log R$ and clearly $\log X$ is not Normal, therefore $X$ is \textbf{not LogNormal}. But it is still valid to calculate $\mathbb{GM}(X)=\mathbb{GM}(S)/\mathbb{GM}(R)$ which leads to
\begin{equation}\label{GMXnonLN}
  \mathbb{GM}(X)\,=\,\frac{b_R}{b_S}\,e^{\psi(a_S)-\psi(a_R)}
\end{equation}
\noindent From (\ref{GMXnonLN}) we have that if $a_R=a_S$ and $b_R=b_S$ then $\mathbb{GM}(X)=1.$ For a power study to compare statistical tests for the null hypothesis $\mathcal{H}_0:\mathbb{GM}(X)=1$ we may use, as before, the equation $\mathbb{GM}(X)=1+\beta$ with $\beta>-1$ and the hypothesis of interest would be equivalent to $\mathcal{H}_0:\beta = 0.$ We will consider two cases:
\begin{equation}\label{aEqual}
  \hspace{-2.3cm}\text{If } a_R\,=\,a_S\,:\qquad \mathbb{GM}(X)\,=\,1+\beta\quad\Leftrightarrow\quad b_R\,=\,(1+\beta)b_S
\end{equation}
\begin{equation}\label{bEqual}
  \text{If } b_R\,=\,b_S\,:\qquad \mathbb{GM}(X)\,=\,1+\beta\quad\Leftrightarrow\quad \psi(a_S)\,=\,\log(1+\beta)+\psi(a_R)
\end{equation}
\noindent Since $\mathbb{E}[\log X]=\mathbb{E}[\log S]-\mathbb{E}[\log R]=\psi(a_S)-\psi(a_R)+\log b_R-\log b_S$ does not involve the (possible) dependence between $S$ and $R$ then $\mathbb{GM}(X)$ depends only on the marginal parameters of $R$ and $S.$ That is not the case of the variance:
\begin{eqnarray}
  \mathbb{V}[\log X] &=& \mathbb{V}[\log S]+\mathbb{V}[\log R]-2\mathbb{C}\text{ov}[\log S,\log R] \nonumber \\
	                   &=& \psi_1(a_S) + \psi_1(a_R) - 2\rho\sqrt{\psi_1(a_S)\psi_1(a_R)} \label{VlogX}
\end{eqnarray}
\noindent where $\rho$ is the linear correlation between $\log S$ and $\log R$ and $\psi_1$ is the \textit{trigamma function} defined by $\psi_1(a):=\psi'(a)=d^{\hspace{0.3mm}2}\log\Gamma(a)/da^2.$ Therefore, the dependence between $S$ and $R$ affects the variability of $\log X$ but not its geometric mean $\mathbb{GM}(X),$ and since the hypothesis of interest only involves the geometric mean, without loss of generality we will assume that $S$ and $R$ are independent Gamma random variables, for the purpose of a power comparison study.

\bigskip

\begin{figure}
  \centerline{
  \includegraphics[width=16cm, keepaspectratio]{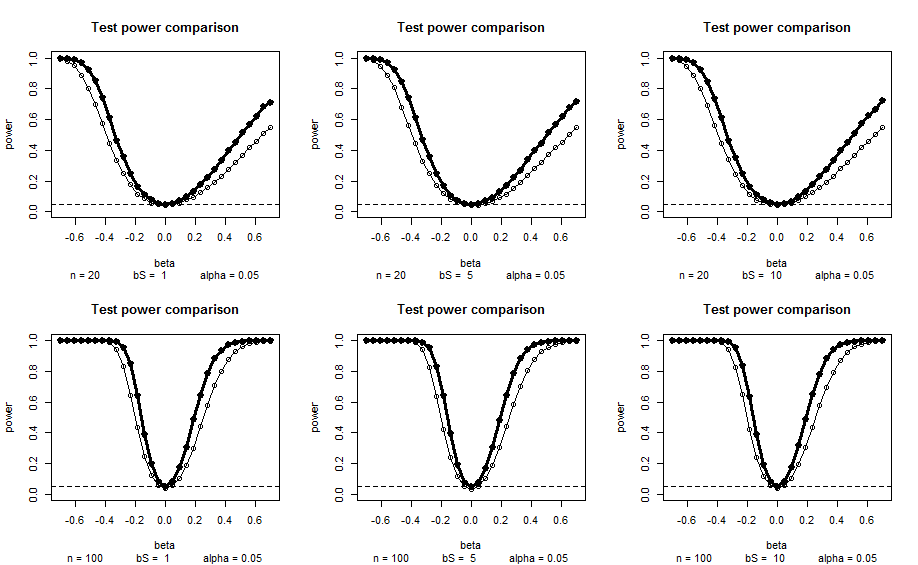}}
  \caption{Estimated power functions for the proposed accuracy test (thick line) versus the binomial test (thin line), for $-0{.}7<\beta<+0{.}7,$ $a_R=a_S=3,$ $b_S\in\{1, 5, 10\}$ and sample sizes $n\in\{20, 100\},$ with $\alpha = 0{.}05$ test level.}
  \label{powerGamma1}
\end{figure}

\noindent The results obtained for the case (\ref{aEqual}) are summarized in Figure \ref{powerGamma1} where it is clear that in all cases the proposed accuracy test (thick line) is uniformly more powerful than the binomial test (thin line). $10,000$ simulations were used to estimate each point of the power functions with $a_R=a_S=3,$ $b_S\in\{1, 5, 10\}$ and sample sizes $n\in\{20, 100\}.$ As expected, the power of both tests increases with a larger sample size, but remains the same independently of $b_S$ and $b_R$ since the variance of $\log X$ does not depend on them, see (\ref{VlogX}).

\bigskip

\begin{figure}
  \centerline{
  \includegraphics[width=16cm, keepaspectratio]{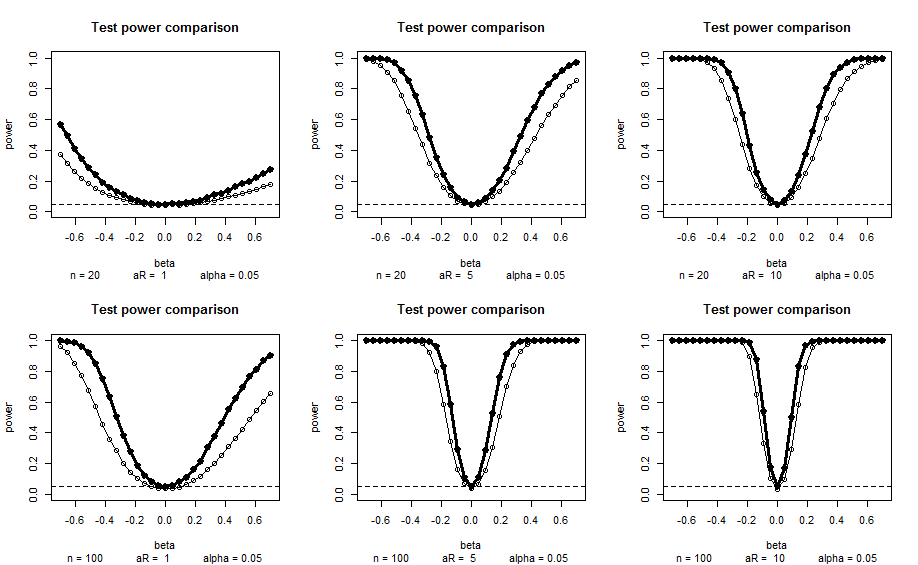}}
  \caption{Estimated power functions for the proposed accuracy test (thick line) versus the binomial test (thin line), for $-0{.}7<\beta<+0{.}7,$ $b_R=b_S=3,$ $a_R\in\{1, 5, 10\}$ and sample sizes $n\in\{20, 100\},$ with $\alpha = 0{.}05$ test level.}
  \label{powerGamma2}
\end{figure}

\noindent The results obtained for the case (\ref{bEqual}) are summarized in Figure \ref{powerGamma2} where it is clear that in all cases the proposed accuracy test (thick line) is again uniformly more powerful than the binomial test (thin line). $10,000$ simulations were used to estimate each point of the power functions with $b_R=b_S=3,$ $a_R\in\{1, 5, 10\}$ and sample sizes $n\in\{20, 100\}.$ As expected, the power of both tests increases with a larger sample size, but now their power is affected by the values of $a_R$ and $a_S$ since the variance of $\log X$ does depend on them, see (\ref{VlogX}).

\bigskip

\noindent Eventhough the log-ratio of two independent Gamma random variables is not normally distributed, the proposed accuracy test had a better power performance than the binomial test. In Table \ref{ShapiroGamma} we show the rejection rate (\% of times p-value $\leq 0{.}05$ in $100,000$ simulations) by the Shapiro-Wilk normality test for a ratio of two independent Gamma$(a,b)$ random variables under sample sizes $n\in\{20,100\}$ with the different $a$ and $b$ values used in the test power comparisons in Figures \ref{powerGamma1} and \ref{powerGamma2}. As a reference, the rejection rate even if the sample is standard normal would be 5\%, and as expected, all the rejection rates are above $5\%$ but in some cases for not so much, which means that these Gamma log-ratios are not too far from being normally distributed.

\begin{table}
\centering
\begin{tabular}{ccccc}
  \hline
 case & a & b & n & \% reject \\ 
  \hline
1 & 3 & 1 & 20 & 6.94 \\ 
  2 & 3 & 5 & 20 & 7.07 \\ 
  3 & 3 & 10 & 20 & 6.89 \\ 
  4 & 3 & 1 & 100 & 10.98 \\ 
  5 & 3 & 5 & 100 & 10.98 \\ 
  6 & 3 & 10 & 100 & 10.99 \\ 
  7 & 1 & 3 & 20 & 11.65 \\ 
  8 & 5 & 3 & 20 & 6.17 \\ 
  9 & 10 & 3 & 20 & 5.49 \\ 
  10 & 1 & 3 & 100 & 30.33 \\ 
  11 & 5 & 3 & 100 & 8.00 \\ 
  12 & 10 & 3 & 100 & 6.26 \\ 
   \hline
\end{tabular}
\caption{12 different combinations of parameters $(a,b)$ for a ratio of two independent gamma random variables, and sample sizes $n\in\{20, 100\}.$ Column \% reject is the percentage of times the Shapiro-Wilk normality test reported a p-value $\leq 0{.}05$ in $100,000$ simulations.} \label{ShapiroGamma}
\end{table}

\section{Application to claims reserving}

According to Kotz \textit{et al.}(2006):
\begin{quote}
Although less obvious as a measure of location than the arithmetic mean, the \textbf{geometric mean} does arise naturally as a measure of
location in at least three circumstances: when observations $X_i$ possess a certain relation between their conditional means and variances, when observed values are thought to be the results of many minor multiplicative (rather than additive) random effects, and when products of moderate to large numbers of random variables are of interest [\ldots] The second circumstance when the geometric mean is relevant occurs when X is the cumulative result of many minor influences which combine in a multiplicative way, so
that the same influence has a greater absolute effect on a larger result than on a smaller one. Since $\log X$ is thus the sum of a great many small random effects, the central limit theorem suggests that $X$ may well be close to log-normal in distribution, even if the contributing influences are not all independent.
\end{quote}

\noindent The above interpretation fits well when the objective is to compare claim amounts paid by an insurance company along several periods ($s_1,\ldots,s_n$), and the forecasted claims for such periods $(r_1,\ldots,r_n).$ To determine if certain actuarial claims reserving method may or may not be considered as accurate, more important than the absolute differences $s_i-r_i$ is to measure how large is that difference as a percentage of what was forecasted, that is $s_i/r_i=(1+\gamma_i)$ means that $s_i$ happened to be $100\gamma_i\%$ above (if $\gamma_i>0$) or below (if $\gamma_i<0$) forecast $r_i\,.$

\bigskip

\noindent Both the proposed accuracy test and the binomial test were applied to the data analyzed in Aguilar and Avenda\~no (2009) as ``Modelo A'', where the authors made a mistake in applying the binomial test by calculating the \textit{p-value} as $\mathbb{P}(B=b)$ instead of the two-sided cumulative probability of the tails, as it should be in two-sided statistical tests. 

\bigskip

\begin{figure}
  \centerline{
  \includegraphics[width=11cm, keepaspectratio]{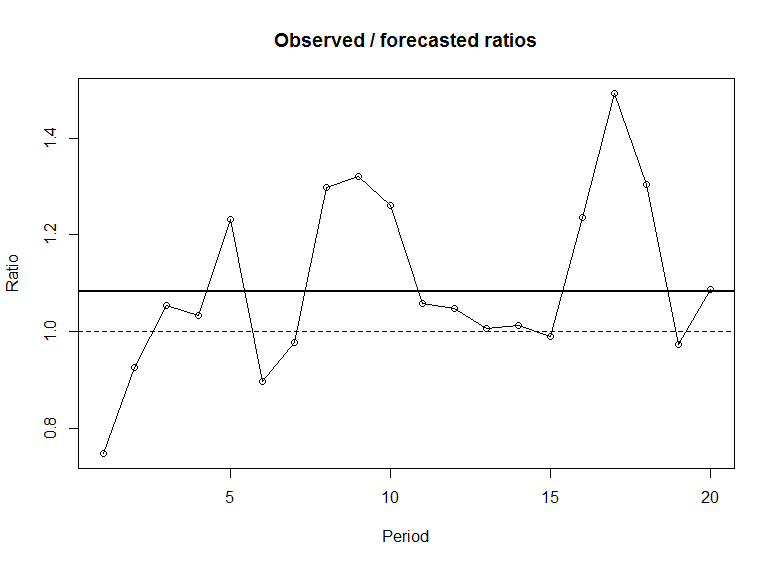}}
  \caption{Ratios of observed/forecasted values of ``Modelo A'' in Aguilar and Avendano (2009). The horizontal thick line level is the value of the sample geometric mean $\widetilde{x}=1{.}083604.$}
  \label{Pedro}
\end{figure}

\noindent The ratios $x_i=s_i/r_i$ for $i=1,\ldots,20$ are shown in Fig.\ref{Pedro} and the horizontal thick line level is the sample geometric mean $\widetilde{x}=(\prod_{i\,=\,1}^{20} x_i)^{1/20}=1{.}083604.$ The proposed accuracy test gives a p-value of $0{.}04092635$ with a Shapiro-Wilk normality p-value of $0{.}5280804$ (so lognormality assumption for ratios is not rejected), while the binomial test gives a p-value of $0.1153183,$ therefore for a significance level $\alpha=0{.}05$ the proposed accuracy test rejects that the forecast method is accurate, but the binomial test fails to reject. Because the lognormality assumption for the ratios is not rejected, we prefer the conclusion from the proposed accuracy test since it is based on a sufficient statistic for the parameter of interest (the geometric mean).

\section{Final remarks}

\noindent In case the lognormality assumption of the ratios observed versus forecasted is not rejected by the Shapiro-Wilk normality test applied to the log-ratios, the proposed accuracy test shows a better power performance than the binomial test, and since it is based on a sufficient statistic for the hypothesis of interest it would be expected to have a good performance compared to some other test. 

\bigskip

\noindent In case the normality of the log-ratios is rejected, there is still the possibility of trying with the more general Box-Cox (1964) transformation $(x_i^{\hspace{0.3mm}\lambda}-1)/\lambda$ for some $\lambda\neq 0$ (recall that the limit of this transformation when $\lambda\rightarrow 0$ is $\log x_i$). 

\bigskip

\noindent An example where the ratios are not lognormal was analyzed (a ratio of Gamma random variables) and the power of the proposed accuracy test seems to be still better than the binomial test, but a more thorough analysis could be made in future work in order to assess the robustness of the proposed test under some other kind of departures from lognormality. But as stated in Johnson \textit{et al.}(1994), even without the lognormality assumption, if $X_1,\ldots,X_n$ are independent and continuous positive random variables, and if the independent random variables $Y_i=\log X_i$ are such that a central limit type of result applies, then the standardized distribution of $\log\widetilde{X}=\frac{1}{n}\sum_{i\,=\,1}^n\log X_i$ would tend to a standard Normal distribution as $n$ tends to infinity, and the limiting distribution of the sample geometric mean $\widetilde{X}$ would then be LogNormal, and the proposed accuracy test may be applied.

\section*{References}

\noindent P. Aguilar-Beltr\'an, J. Avenda\~no-Estrada (2009) \textit{Backtesting Insurance Reserves}, Mexican Association of Insurance Companies (AMIS). Document retrieved on September 16th 2015 at the following internet link http://www.amis.org.mx/InformaWeb/Documentos/Archivos\newline/Articulo\_\%20Backtesting\_Septiembre\_2009.pdf \newline Permanent link https://goo.gl/9PxE82

\medskip

\noindent G.E.P. Box, D.R. Cox (1964) An analysis of transformations. \textit{J. Royal Statist. Soc. B} \textbf{26} (2), 211--252.

\medskip

\noindent F.X. Diebold, R.S. Mariano (2002) Comparing Predictive Accuracy. \textit{Journal of Business \& Economic Statistics} \textbf{20}(1), 134--144.

\medskip	

\noindent R.J. Hyndman, A.B. Koehler (2006) Another look at measures of forecasting accuracy. \textit{International Journal of Forecasting} \textbf{22}, 679--688.

\medskip

\noindent N.L. Johnson, S. Kotz, N. Balakrishnan (1994) \textit{Continuous univariate distributions, Volume 1}, Wiley, New York.

\medskip

\noindent S. Kotz, N. Balakrishnan, C.B. Read, B. Vidakovic, N.L. Johnson (2006) \textit{Encyclopedia of Statistical Sciences}, Wiley, New Jersey.

\medskip

\noindent A.J. McNeil, R. Frey, P. Embrechts (2015) \textit{Quantitative Risk Management}, Princeton University Press, New Jersey.		

\medskip

\noindent A.M. Mood, F.A. Graybill, D.C. Boes (1974) \textit{Introduction to the Theory of Statistics}, McGraw-Hill, New York.

\medskip

\noindent N.M. Razali, Y.B. Wah (2011) Power comparisons of Shapiro-Wilk, Kolmogorov-Smirnov, Lilliefors, and Anderson-Darling tests. \textit{Journal of Statistical Modeling and Analytics} \textbf{2} (1), 21--33.

\medskip	

\noindent S.S. Shapiro, M.B. Wilk (1965) An Analysis of Variance Test for Normality. \textit{Biometrika} \textbf{52} (3-4), 591--611.

\medskip

\noindent M.V. Shcherbakov, A. Brebels, N.L. Shcherbakova, A.P. Tyukov, T.A. Janovsky, V.A. Kamaev (2013) A Survey of Forecast Error Measures. \textit{World Appl. Sci. J.} \textbf{24} (Information Technologies in Modern Industry, Education \& Society), 171--176.
  
\end{document}